\documentstyle[12pt,prl,multicol,aps,epsfig]{revtex}
\def\ltsim{\mathop{\raise3pt\hbox{$<$}\llap{\lower3pt\hbox{$\sim
$}}}}
\def\gtsim{\mathop{\raise3pt\hbox{$>$}\llap{\lower3pt\hbox{$\sim
$}}}}
\newcommand{\be}{\begin{eqnarray}}
\newcommand{\ee}{\end{eqnarray}}

\begin{document}

%\draft
%\tighten
%\preprint{}
%[\hsize\textwidth\columnwidth\hsize\csname @twocolumnfalse\endcsname

\title{Single hole dynamics in the one dimensional $t$-$J$ model}
\author{Michael Brunner, Fakher F. Assaad  and   Alejandro Muramatsu}
\address{
Institut f\"ur Theoretische Physik III, Universit\"at Stuttgart,
Pfaffenwaldring 57, D-70550 Stuttgart,\\ 
Federal Republic of Germany
}
\date{\today}
\maketitle
\begin{abstract}
We present a new finite-temperature quantum Monte Carlo algorithm
to compute imaginary-time Green functions for a single hole in 
the $t$-$J$ model on non-frustrated lattices.
Spectral functions are then obtained
with the Maximum Entropy method. Simulations of the one-dimensional 
case show that a simple charge-spin separation {\em Ansatz}
is able to describe the overall features of the spectral function over 
the whole energy range for values of $J/t$ from $1/3$ to $4$.
This includes the bandwidth  $W \sim 4t + J$  and the compact support
of the spectral function. The quasiparticle weight $Z_k$ is computed on
lattices up to $L=96$ sites, and scales as $Z_k\propto L^{-1/2}$.

\end{abstract}
%]Lattice fermion models, fermions in reduced dimension
%\newpage
%\begin{multicols}{2}
\par
%\section{Introduction}
Understanding single hole dynamics in quantum antiferromagnets 
is a decisive step towards a comprehensive description of elementary
excitations in strongly correlated systems. Experimental realizations 
are found in compounds such as $SrCuO_2$ \cite{kim96}, $Na_{0.96}V_2O_5$ 
\cite{kobayashi99} for chains, 
$Sr_{14}Cu_{24}O_{41}$  \cite{takahashi97} for ladders and
$Sr_2 Cu O_2 Cl_2$ \cite{wells95} for planes. In particular chain compounds attract 
at present an increasing amount of interest in order to elucidate, 
whether signals of charge-spin separation as predicted from 
Luttinger-liquid theory can be observed experimentally. On the other hand, 
theoretical treatments based on Bethe-{\em Ansatz} (BA) results lead recently to 
a complete description of the spectral function of the Hubbard model
at $U=\infty$
\cite{sorella92}
%,penc95} 
and the low energy sector in the 
nearest-neighbour (NN) $t$-$J$ model, where explicit results are obtained
at the supersymmetric (SuSy) point 
\cite{sorella96}.
%,sorella98}.
Further exact results - apart from exact
diagonalizations 
which suffer from strong finite-size effects - are available only for the 
inverse-square exchange (ISE) 
\cite{kato98}
%,ha94,haldane94}
$t$-$J$ model at the SuSy point.
In order to be able to compare with experiments, it is crucial to extend 
such studies to realistic values of the parameters and possibly beyond 
the asymptotic low energy limit.

In this letter, we present a simple finite-temperature quantum 
Monte Carlo (QMC) algorithm capable of dealing with this issue 
for the NN $t$-$J$ model. For single-hole excitations and in the 
absence of 
frustration, the method is free of the notorious sign problem, 
and applicable to chains, n-leg ladders and planes. 
Here, we concentrate on chains. Our simulations lead to 
the conclusion that the overall features of the spectral functions are 
well described by a charge-spin separation {\em Ansatz} (CSSA) based on
a mean-field slave-boson picture \cite{suzuura97},
where the hole spectral function is given by the convolution of the
spectral functions of free holons and spinons. The agreement with the 
simulations is obtained over all energy scales and values of $J/t$ ranging 
from $1/3$ to $4$. At the SuSy point a more detailed understanding of the
spectrum is achieved by supplementing the simple model with 
BA results.
%Bethe-{\em Ansatz}
A finite-size scaling on chains
up to $L=96$ sites 
shows that the quasi-particle weight $Z_k$ vanishes as $1/\sqrt{L}$,
a result
which was beyond numerical capabilities up to now.
 
Our starting point is the NN $t$-$J$ model,
%with next neighbour interaction only,
\be
H_{t-J}=
-t \sum\limits_{<i,j>,\sigma} \tilde c^{\dagger}_{i,\sigma} \tilde 
c^{}_{j,\sigma} 
+J \sum\limits_{<i,j>}
\left( \vec S_i\cdot \vec S_j -\frac 1 4 \tilde n_i \tilde n_j \right).
\ee
Here $\tilde c^{\dagger}_{i,\sigma}$ are projected fermion operators
$\tilde c^{\dagger}_{i,\sigma}=(1-c^{\dagger}_{i,-\sigma}
c^{}_{i,-\sigma})c^{\dagger}_{i,\sigma}$
,
$\tilde n_i=\sum\limits_{\alpha} \tilde c^{\dagger}_{i,\alpha}\tilde
c^{}_{i,\alpha}$,
$\vec S_i=(1/2)\sum\limits_{\alpha,\beta}c^{\dagger}_{i,\alpha}
\vec{\sigma}_{\alpha,\beta}c^{}_{i,\beta}$,
and the sum runs over nearest neighbours.
After a canonical transformation this model
is cast into the form 
\cite{khaliullin90}
%,antimo95}:
\be
\tilde H_{t-J}= +t \sum\limits_{<i,j>} P_{ij}f^{\dagger}_if^{}_j + 
\frac J 2 \sum\limits_{<i,j>}\Delta_{ij}(P_{ij}-1) ,
\ee
where
$P_{ij}=(1+\vec\sigma_i\cdot\vec\sigma_j)/2$,
$\Delta_{ij}=(1-n_i-n_j)$ and
$n_i=f^{\dagger}_if^{}_i$.
In this mapping, one uses the following identities for the 
standard creation ($c^{\dagger}_{i,\sigma}$) and annihilation
($c^{}_{i,\sigma}$) operators 
%for fermions with spin $\sigma= \uparrow , \downarrow$:
%
%\begin{eqnarray}
%\label{newrep}
%
$
c^\dagger_{i\uparrow} = \gamma_{i,+} f_i - \gamma_{i,-} f_i^\dagger \, , 
\; \; \;
c^\dagger_{i\downarrow} = \sigma_{i,-} (f_i + f_i^\dagger) \, ,
$
%\end{eqnarray}
where $\gamma_{i,\pm} = (1 \pm \sigma_{i,z})/2$ and $\sigma_{i,\pm} =
(\sigma_{i,x} \pm i \sigma_{i,y})/2$. The spinless fermion operators 
fulfill the canonical anticommutation relations 
$\{f_i^\dagger,f_j\} = \delta_{i,j}$,
and $\sigma_{i,a}\, , \; a = x,y,$ or $z$ are the Pauli matrices. 
The constraint to avoid doubly occupied states transforms to the
conserved and holonomic constraint
$ \sum_i \gamma_{i,-} f^\dagger_i f_i = 0 $.

The Green function in the spin up sector may  be written as
\be
G_{\uparrow}(i-j, \tau)=
\langle T \tilde c^{}_{i,\uparrow}(\tau) \tilde c^{\dagger}_{j,\uparrow}
\rangle
= 
\langle 
 T f^{\dagger}_i(\tau) f^{}_j 
\rangle
\ee
where $T$ corresponds to the time ordering operator.
Inserting complete sets of spin states the quantity above transforms as
%\end{multicols}
\begin{eqnarray}
-G(i-j,-\tau) 
 = & &
\frac{\sum \limits_{\sigma_1} 
\langle v | \otimes \langle \sigma_1 | e^{-(\beta-\tau)\tilde H_{t-J}}f^{}_j
e^{-\tau \tilde H_{t-J}}f^{\dagger}_i | \sigma_1 \rangle \otimes | v \rangle}
{\sum \limits_{\sigma_1}\langle \sigma_1 |
 e^{-\beta\tilde H_{t-J}} | \sigma_1 \rangle} 
= \nonumber \\ 
& &
\sum\limits_{\vec\sigma} P(\vec\sigma) \times 
\frac{
\langle v|  f^{}_j 
e^{-\Delta\tau\tilde H(\sigma_n,\sigma_{n-1})}
e^{-\Delta\tau\tilde H(\sigma_{n-1},\sigma_{n-2})}
\ldots
 e^{-\Delta\tau\tilde H(\sigma_{2},\sigma_{1})}
 f^{\dagger}_i
|v \rangle
}
{
\langle\sigma_n |  e^{-\Delta\tau\tilde H_{t-J}} | \sigma_{n-1}\rangle
\ldots \langle \sigma_{2}  e^{-\Delta\tau\tilde H_{t-J}}
| \sigma_{1}\rangle 
}
+{\cal O}(\Delta\tau^2)
\nonumber \\
& & = 
\sum\limits_{\vec\sigma} P(\vec\sigma) G(i,j,\tau,\vec\sigma)
+{\cal O}(\Delta\tau^2)
%\end{multline}
\end{eqnarray}
%\begin{multicols}{2}
Here $m \Delta\tau=\beta$,   $  n \Delta\tau=\tau$, $\Delta\tau t \ll 1$
and $\exp({-\Delta\tau\tilde H(\sigma_1,\sigma_2)})$ is the
evolution operator for the holes, 
given the spin configuration $(\sigma_1,\sigma_2)$.
In the case of single hole dynamics 
$|v\rangle$ is the vacuum state for  holes, and
$P(\vec\sigma)$ is the probability distribution of 
a Heisenberg antiferromagnet for
the configuration $\vec\sigma$, where
$\vec \sigma$ is a vector containing all intermediate states
$(\sigma_1,\ldots\sigma_n,\ldots\sigma_{m},\sigma_1)$.
The sum over spins is performed in a very efficient way 
by using a world-line
cluster-algorithm
% with discretized imaginary time
%\cite{evertzsuzuki,kawashima94}.
\cite{evertz93}.
As the evolution operator for the holes is a bilinear form
in the fermion operators, 
$G(x,\tau,\vec\sigma)$
can be calculated exactly.
$G(x,\tau,\vec\sigma)$ contains a sum over all possible fermion paths
between $(i,0)$ and $(j,\tau)$, where $i-j=x$. This stands in contrast
to the worm approach \cite{prokofev98}, where fermion paths are sampled 
stochastically.
The numerical effort to calculate
$G(x,\tau,\vec\sigma) \, \forall \, x,\tau$  scales as $L \tau$.
Spectral properties are obtained by inverting the spectral theorem
\be
G (k,\tau)=\int\limits_{-\infty}^{\infty} {\rm d}\omega
A(k,\omega)\frac{
\exp(-\tau\omega)}{\pi(1+\exp(-\beta\omega))} 
\ee
%\be A(k,\omega)= 1/N \sum\limits_{N_p,n,m} 
%\frac{\exp(-\beta E_n)}{Z}
%\left|
%\langle \Psi_m^{N_p-1}|c^{}_{k\sigma}|\Psi_n^{N_p}\rangle\right|^2
%\delta(\omega-E_n^{N_p}+E_m^{N_p-1})
%\ee
with the Maximum Entropy method (MEM)
\cite{jarrell96}.
%,linden95}.
%With the representation (\ref{newrep}), the propagation of down spin 
%electrons cannot be easily considered,
%since the operators $\sigma_{i,\pm}$ cut world-lines.
%This is certainly not a problem for finite-size systems,
%where SU(2) symmetry is conserved. 
Since $P(\vec\sigma)$ is the probability distribution 
for the quantum antiferromagnet, the algorithm does not suffer from 
sign problems
on bipartite lattices and next neighbour interactions in any dimension.
However, when the spin and charge dynamics evolve according to very
different time scales ($J \ltsim 0.2 t$), $G(x,\tau,\vec\sigma)$ shows an 
increasing variance. Best results are obtained at the SuSy point and
an appreciable range of $J/t$ may be considered as shown below.

We now concentrate on the one-dimensional $t$-$J$ model.
The simulations were performed at temperatures $T \le {\rm min} (J,t) /15$, 
such that no appreciable changes with a further decrease in temperature 
can be seen: the results correspond to the zero temperature limit, 
a limit which is in general difficult to reach in other finite-temperature
fermionic algorithms. 
We compare our results with the predictions of the CSSA, 
where free holons and spinons are described by
\cite{suzuura97,anderson87}
%,baskaran88}
\be
\label{sbos}
H=-\frac {t_h} 2 \sum\limits_{<i,j>}h^{\dagger}_ih^{}_j 
- \frac {J_s} 2 \sum\limits_{<i,j>}s^{\dagger}_{i,\sigma}s^{}_{j,\sigma}.
\ee
Here the electron operator $c^{}_{i,\sigma}$ is given by the product 
of a holon ($h_i$) and a spinon ($s_{i,\sigma}$) operator,
$c^{}_{i\sigma}=s_{i,\sigma}h^{\dagger}_i$,
the holon being a boson and the spinon a spin-1/2 fermion. 
As a consequence of the above {\em Ansatz} , the dispersion relations of 
the free holons and spinons are given by $\epsilon_h = - t_h \cos q_h$
and $\epsilon_s = - J_s \cos q_s$ respectively, whereas the energy of the
hole is $E(k) = \epsilon_h - \epsilon_s$ and by momentum conservation
$k = q_h - q_s$. We take
$t_h$ and $J_s$ as two free parameters in contrast to a mean-field 
approximation, where they have to be calculated self-consistently. 
The spectral function is then given by a convolution of the spinon
and holon Green functions. The lowest attainable energy ($-t_h$)
and highest one ($t_h + J_s$) define the bandwidth of the hole, 
$2 t_h + J_s$.
Since the full band-width obtained by considering the compact support
of the spectral function at $J=0$ is known to be exactly $4t$ 
\cite{sorella92}, we take $t_h = 2t$. In order to determine $J_s$, 
we consider the overall bandwidth, as obtained from the simulation.
As can be seen in Fig.\ 1, for all values of $J$,
\begin{figure}[bth]
\centering
\epsfig{file=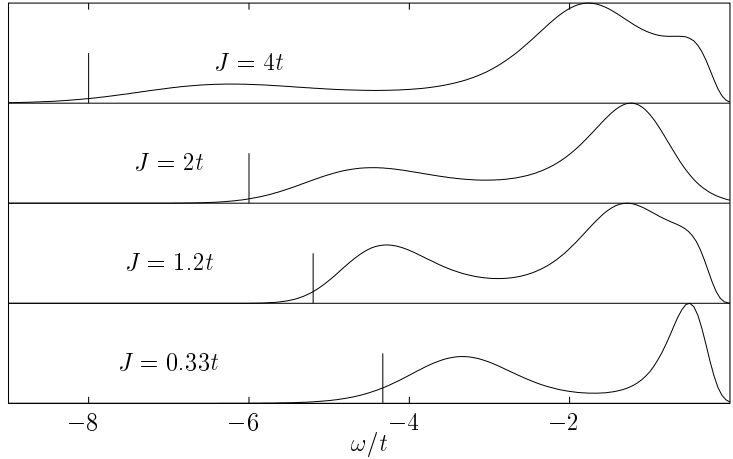,width=11cm}
\caption{Density of states $N(\omega)$ for different values of $J/t$.
The vertical line
indicates $4t+J$.}
\end{figure}
the width of the density of states $N(\omega)$
scales approximately as $4t+J$ in the parameter range considered,
leading to $J_s = J$.

Beyond predicting bandwidths, the CSSA describes
accurately the support of the spectral function in the case $J=0$,
when compared with exact results \cite{sorella92,suzuura97}.
If furthermore phase string effects \cite{suzuura97} are taken into account,
the singularities of $A(k,\omega)$ related to holons and spinons
can be reproduced. For finite $J$, the minimal (maximal) possible energy 
of a hole in 
CSSA is given by $E(k) = - F_k$ ($E(k) = F_k$) for $k < k_0$ ($k> k_0$),
where $F_k \equiv \sqrt{J^2+ 4t^2-4tJ\cos\left(k\right)}$
contains both holon and spinon contributions, and $k_0$ is determined by
$\cos(k_0)=J/(2t)$. The remaining parts of the compact support are given
by $E(k) = \mp 2t\sin (k)$ for $k > k_0$ (lower edge)
and $k<k_0$ (upper edge)
respectively. Such dispersions correspond to holons with momentum $k + q_s$,
and a spinon with $q_s = \mp \pi/2$ \cite{sorella96,suzuura97}.
As $J \rightarrow 2t$, 
$k_0 \rightarrow 0$ and the lower edge of the compact support is entirely 
determined by the dispersion of the holon. 

\begin{figure}[bth]
\centering
\epsfig{file=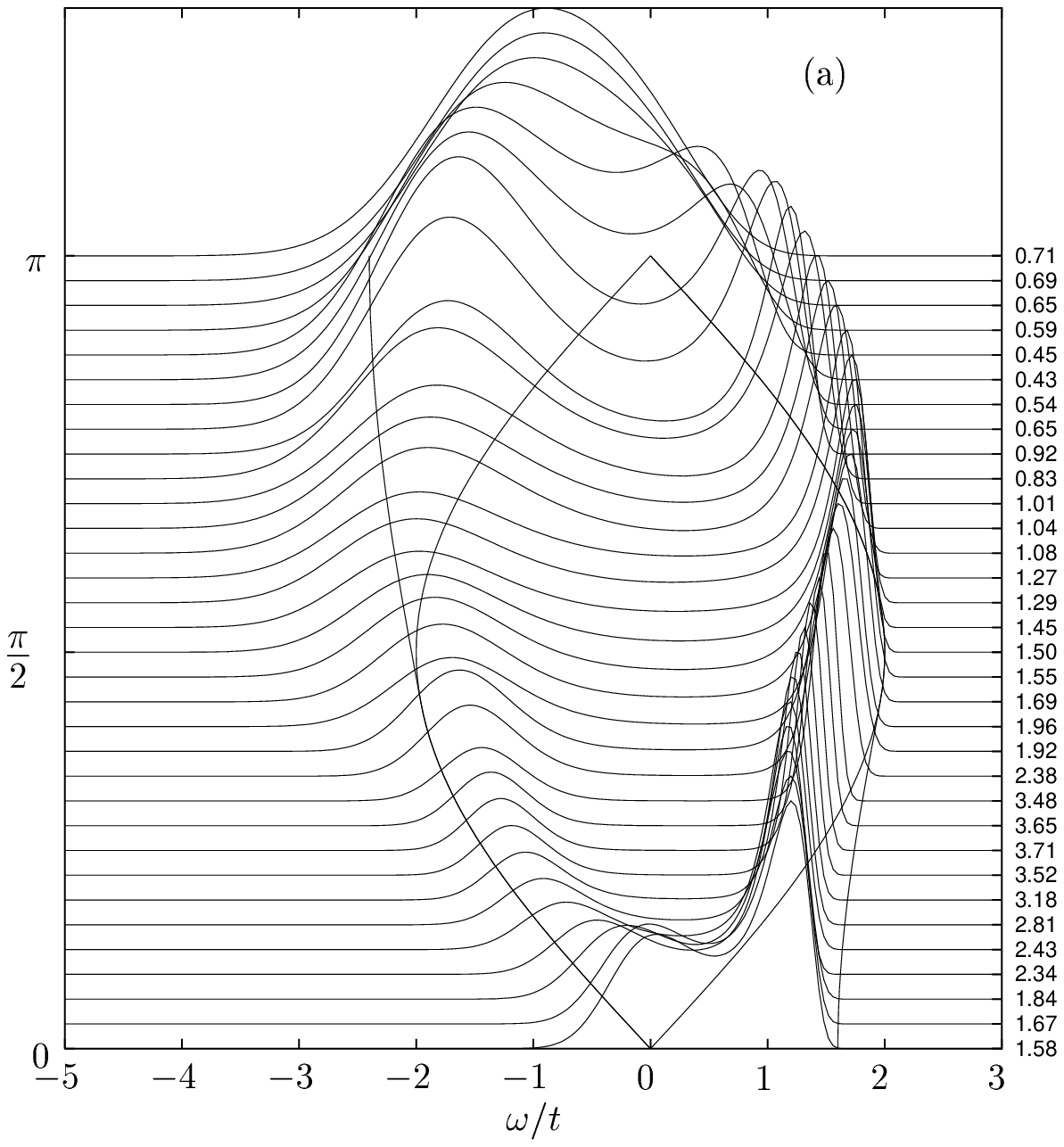,width=8.1cm}
\epsfig{file=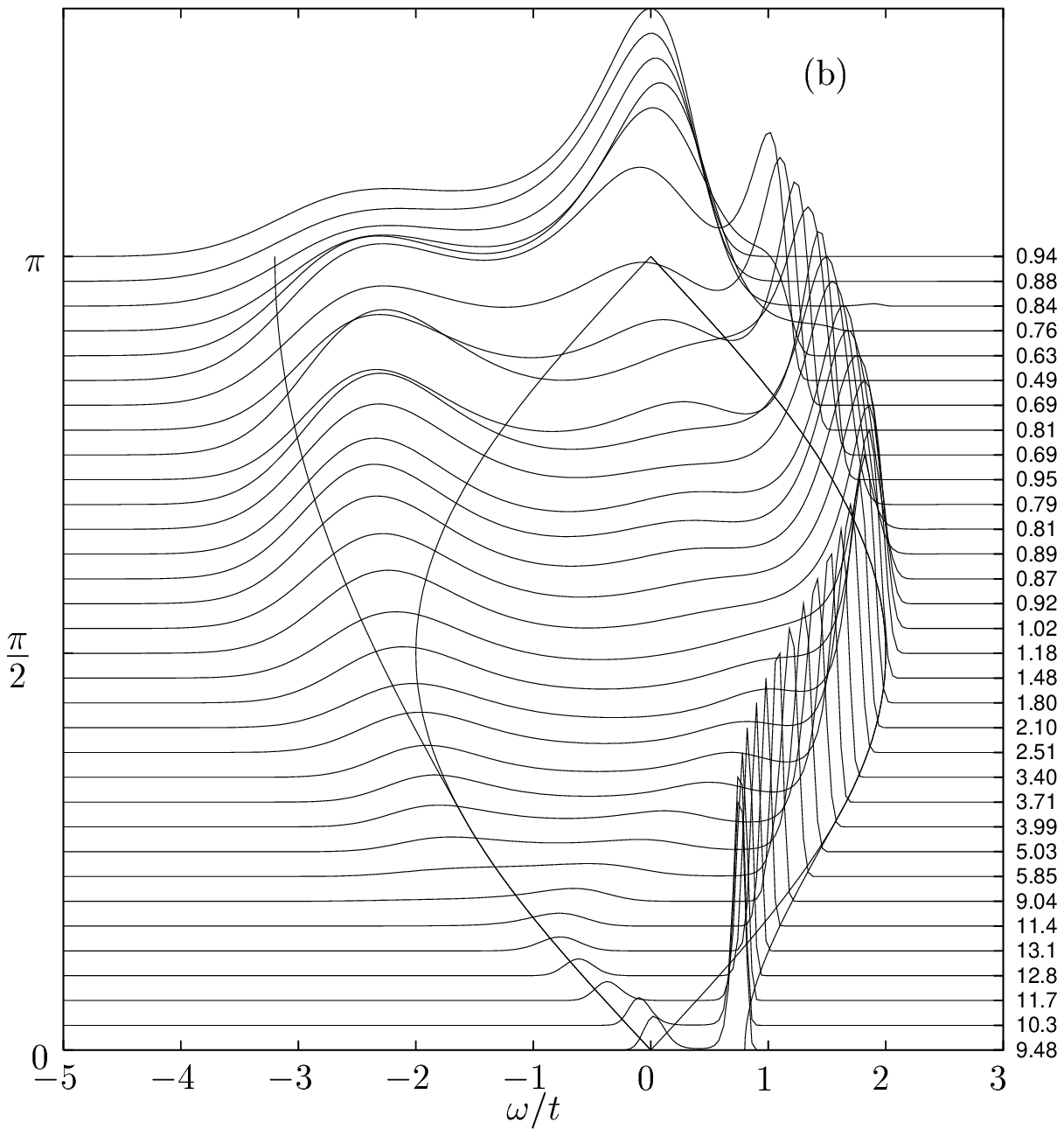,width=8.1cm}
\epsfig{file=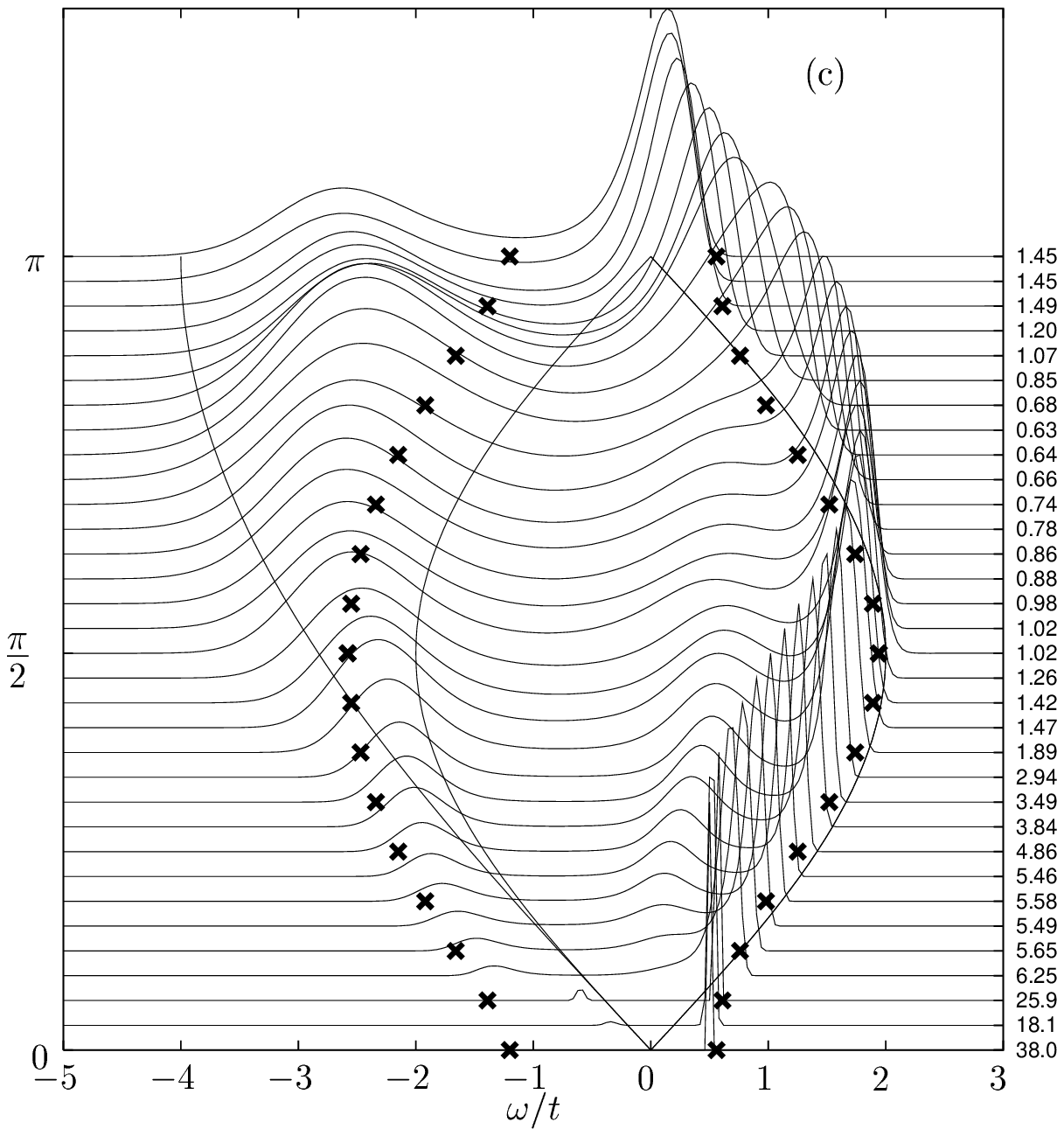,width=8.1cm}
\caption{Spectral function $A(k,\omega)$ for $J=0.4t$ (a),
$J=1.2t$ (b) and $J=2t$ (c).
Here the wave vector $k\in [0,\pi]$ is given on the y-axis.
For clarity, the data  is rescaled by the
number given at the right hand side
of the plot.
Further details are discussed in the text.
%
%The full line indicates the result of the CSSA,
%the crosses for $J/t=2$
%give the lower and high energy cutoff for a
%noninteracting spinon-holon pair, with dispersions given
%by the Bethe-{\em{}Ansatz} [15].
%\cite{bares90,bares91}.
}
\end{figure}
We now compare the above predictions with our QMC data. Figure 2 shows
$A(k,\omega)$ for $J/t=0.4$ (a), 
$1.2$ (b) and $2$ (c).
In all cases the compact support is reproduced very well 
by the CSSA. The {\em Ansatz} also predicts
singularities at the lower (upper)
edge for $k < k_0$ ($k> k_0$), and when phase strings are considered 
\cite{suzuura97} along the edges and the holon lines ($\pm 2t\sin (k)$) for
all momenta. The singularities along the lower holon
line are also supported by
a recent low energy theory \cite{sorella96}. For all parameter values
we observe
dominant weight along the above mentioned lines. For $J/t=0.4$,
we have checked that the results are consistent within the uncertainties of 
MEM with a peak along the edges and a further peak along the holon lines, 
signaled by a broad structure between the edges and the holon
lines (Fig.\ 2.a). We observed such a behaviour for 
$0.33 \leq J/t \leq 0.6$. For $J/t \geq 1.2$ (Fig.\ 2.b and 2.c),
the structure at the lower edge
narrows considerably and the data are not any more consistent with an
additional structure along the lower holon line
for $k < k_0$, but only with
a singularity for $k > k_0$. At $J/t=2$ the exact holon and spinon
dispersions can be obtained by BA
%Bethe-{\em Ansatz} 
\cite{bares90}.
%,bares91}.
Figure 2.c 
shows the comparison with the CSSA, where on the one side
the original dispersions are used (full line) and on the other side,
with the dispersions as given by 
%Bethe-{\em Ansatz}
BA (crosses).  
Whereas the BA
%Bethe-{\em Ansatz}
holon dispersion reproduces very well the lower edge, showing
that as anticipated by the CSSA, at the SuSy point that edge
is completely determined
by the holon dispersion, the full bandwidth is better described with the 
original dispersions. We assign the additional weight in the region
$k > \pi/2$ to processes involving one holon and more than one 
%(Bethe-{\em Ansatz}) 
BA spinon. In fact, that portion resembles the difference
between the supports for one-holon/one-spinon and one-holon/three-spinon
processes in the ISE model
\cite{kato98}.
%,ha94}.
In our case, no limitation
on the possible number of spinons exists, such that in principle all odd
number of them are allowed. It is interesting to notice that using a 
fermionic spinon one is able to describe both the case $J=0$ and $J=2t$.
In the first case, the spinon in the exact solution is a fermion. At the
SuSy point it is expected to be a semion 
\cite{kato98,haldane94}
%haldane94,ha94}
and on the basis
of our results, we conclude that the fermionic spinon contains all possible 
states with an odd number of semionic spinons.
\par
Finally, we consider the quasiparticle residue
$ Z_k = \left| \langle  \Psi_0^{L-1} | 
\tilde c^{}_{k\sigma} | \Psi_0^{L} \rangle\right|^2$ at $k = \pi/2$ for 
$J=2t$.
\begin{figure}[bth]
\centering
\epsfig{file=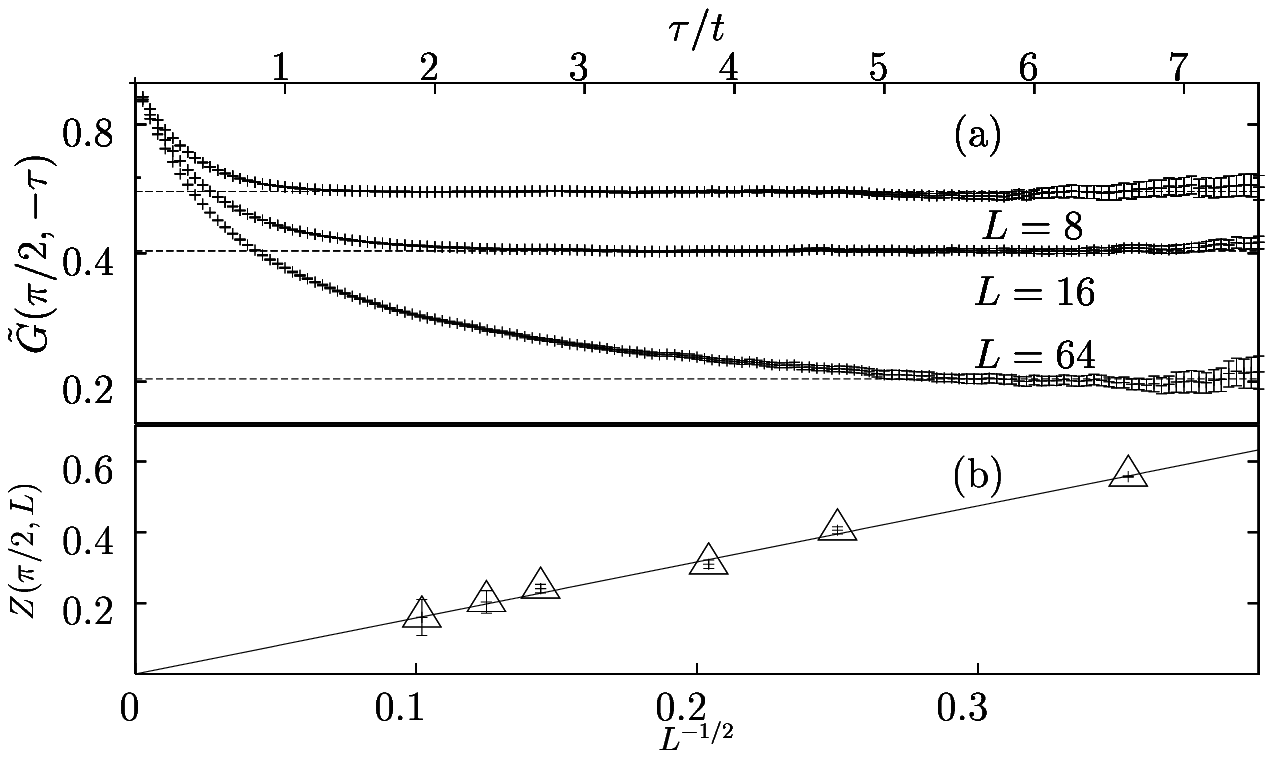,width=11cm}
\caption{Quasiparticle weight at $k=\pi/2$.
\hfill $\mbox{ }$ (a)
$\tilde G(\pi/2,-\tau)\equiv
G(\pi/2,-\tau) \exp\left[-\tau\left(
E_0^L-E_0^{L-1}\left(\pi/2\right)\right)\right] $
versus $\tau/t$.
At $\tau/t \gg 1$ this quantity converges
to the quasiparticle weight $Z(\pi/2)$.
\hfill $\mbox{ }$ (b)
Finite size scaling of $Z(\pi/2)$ as obtained from (a).
The solid line is
a least square fit to the form $L^{-1/2}$.
We consider $\beta J =30$ for $L\le 48$ and $\beta J=60$ for $L>48$,
to guarantee convergence in $\tau$.}
\end{figure}
As Fig.\ 2.c shows, the lower edge is very sharp and without prior
knowledge, the question may arise whether
we are dealing with a quasiparticle.
$ Z_k$ is related to the imaginary time Green function through:
\be
\lim_{\tau \rightarrow \infty}  G(k, -\tau) \propto Z_k 
\exp \left[ \tau \left(E_0^{L} - E_0^{L-1}\left(k\right) \right) \right]
.
\ee
Fig.\ 3.a shows 
$G(\pi/2, -\tau) \exp \left( - \tau (E_0^{L} - E_0^{L-1}(\pi/2) ) \right)$
versus $\tau$, where the energy difference is obtained by fitting
the tail of $G(\pi/2, -\tau)$ to a single exponential form, 
for several sizes. The 
thus estimated $ Z(\pi/2) $ is plotted versus system size in Fig.\ 3.b. Our 
results are consistent with a vanishing quasiparticle weight
$Z(\pi/2)  \propto L^{-1/2}$ which is
the scaling obtained by a combination of bosonization and conformal field 
theory \cite{sorella96}. Since the CPU-times scales as $V\beta$ ($V$ is 
the volume) the determination of the $Z$-factor may be efficiently extended 
to higher dimensions, in contrast to determinantal
algorithms for the Hubbard 
model that scale as $V^3\beta$.

In summary, we have developed a new QMC algorithm which allows the 
determination of single-hole dynamics in quantum antiferromagnets.
This algorithm is extremely powerful in the sense, that the required CPU time
scales
as $V\beta$. For the one dimensional
case, we showed that the spectral function is well described by a simple model
with free spinons and holons with dispersions given by $J$ and $2t$
respectively. The comparison of our results at the supersymmetric
point lead to a characterization of the excitation content of the spectra 
for this particular parameter, where additional information is available
from the Bethe {\em Ansatz} solution. Finally 
we computed the quasiparticle weight and showed that it vanishes as
$L^{-1/2}$.

This work was supported by Sonderforschungsbereich 341 in
T\"ubingen-Stuttgart. The numerical calculations were performed at
HLRS Stuttgart and HLRZ J\"ulich. We thank the above institutions for
their support.

%\bibliographystyle{/user/sams/people/michi/tex/texinput/prsty}
%\bibliography{/user/sams/people/michi/tex/texinput/setenza}

%\end{multicols}
\end{document}